\documentclass[reprint,
superscriptaddress,
 amsmath,amssymb,
 aps,
pre,
]{revtex4-2}

\usepackage{graphicx}
\usepackage{dcolumn}
\usepackage{bm}
\usepackage{xcolor}

\begin{document}

\preprint{APS/123-QED}

\title{Lossy anharmonic polaritons under periodic driving}

\author{Maicol A. Ochoa}
\email{maicol@umd.edu}
\affiliation{Department of Chemistry and Biochemistry, University of Maryland, College Park,MD, USA}


%
%

\date{\today}

\begin{abstract}

 We report on the anharmonic signatures in dissipative polaritons' stationary energy distribution and thermodynamics under external periodic driving. First, we introduce a dynamic model for the dissipative anharmonic Jaynes-Cummings polariton with a generic time-periodic interaction representing modulations of the polariton's energy due to an external force or field. We characterize the stationary state in terms of the exciton, phonon, and interaction energy dependence on the phonon anharmonicity, exciton-phonon coupling strength, and intensity and form of the external field-polariton coupling. Our model also captures the quantum thermodynamics of the driven polariton, which we analyze in connection with the irreversible heat, maximum power, and efficiency of the process. We find considerable differences in energy distribution and thermodynamics between harmonic, moderate, and strongly anharmonic polaritons. Moreover, comparing the external modulations to the phonon and exciton energy, we conclude that the former enhances the polariton's energy storage capacity and is occasionally limited by interference effects and energy saturation at the exciton.

\end{abstract}

\maketitle



\section{Introduction}\label{sec:int}

Polaritons originating on light-matter interactions between quantum emitters and cavities \cite{li2021molecular,dunkelberger2022vibration}, molecules near plasmonic nanoparticles\cite{fofang2011plexciton,bitton2022plasmonic,mondal2022coupling,mondal2022strong}, optically-stimulated semiconductors\cite{kaitouni2006engineering,deng2010exciton,byrnes2014exciton,ochoa2020extracting,khan2021efficient,denning2022cavity}, and nanomechanical devices\cite{pigeau2015observation,li2016hybrid,munoz2018hybrid}, are ideal systems to investigate hybrid quantum state formation, their optical response, design and quantum control\cite{blais2021circuit}. Typically, the exciton-photon coupling strength modulates the polariton characteristics, ranging from regimes with weak to strong coupling relative to the exciton and photon energies. A weakly-coupled composite shows properties that are close to those of its components, while strong coupling often leads to hybrid states with distinct and unique properties\cite{ochoa2016energy}. External fields and forces provide additional control on the polartonic state. This tunability makes polaritions attractive platforms for quantum technologies\cite{kurizki2015quantum,ghosh2020quantum}, quantum materials\cite{guan2022light}, polaritonic chemistry\cite{li2022molecular,mandal2023theoretical,campos2023swinging}, and nanoscale devices.

The Jaynes-Cummings model\cite{shore1993jaynes,casanova2010deep,bina2012coherent} (JCM) is the prime model from which most basic polariton physics has been rationalized, serving also as the foundation for more elaborate representations\cite{larson2022jaynes}, such as the Tavis-Cummings model\cite{fink2009dressed,agarwal2012tavis,zhu2016effects,sun2022dynamics,campos2021generalization}, the quantum Rabi model\cite{xie2017quantum,rossatto2017spectral} and the Dicke model\cite{chen2008numerically,garraway2011dicke}. In the JCM a two-level system and a single phonon represent correspondingly the exciton and the light, and the energy-preserving interactions assumes the rotating-wave approximation. Adding relaxation to the JCM model, either phenomenologically\cite{garraway1997nonperturbative,wang2021light,deffner2014optimal,li2022polariton} or by including explicit reservoirs\cite{barnett1986dissipation,del2015quantum,martinez2018comment,ochoa2022quantum,bleu2023effective}, transforms the JCM into a dissipative open system modified by the environment properties, such as the environment's spectral density and temperature. When the polaritonic state consists of an emitter in a cavity, these dissipative processes account for deviations of the real system from the ideal JCM model.  Recent studies in the polariton dynamics \cite{bamba2012dissipation,sieberer2013dynamical,drezet2017quantizing,pistorius2020quantum} research on the  dissipation\cite{kyriienko2014optomechanics,liu2014coherent,pistorius2020quantum} in open or lossy systems. Moreover, it is imperative to account for environment-polariton interactions and  external driving fields to analyze techonological applications deriving from the polariton's response and dynamics. The field of quantum thermodynamic\cite{pekola2015towards,esposito2015quantum,vinjanampathy2016quantum,ochoa2018quantum,deffner2019quantum,kosloff2019quantum,kurizki2022thermodynamics} develops physically meaningful definitions for quantities such as work\cite{talkner2007fluctuation}, heat\cite{esposito2015nature,pekola2015towards,whitney2018quantum}, entropy\cite{esposito2015quantum,kalaee2021positivity}, and efficiency\cite{esposito2015efficiency} applicable to exciton polaritons in nonequilibrium. Recent studies focus on the dynamics of periodically driven nanoscale systems\cite{ochoa2015pump,chen2023energy} utilizing methods based on reduced quantum master equations\cite{hotz2021coarse,tanimura2020numerically,wacker2022nonresonant,ochoa2022quantum} and Floquet theory\cite{hone2009statistical,restrepo2018quantum,deng2016dynamics,wang2023nonadiabatic,batge2023periodically,wang2024nonadiabatic}.

In Ref.\ \citenum{ochoa2022quantum}, we investigated the quantum thermodynamics of a periodically driven polariton after developing a consistent dynamic and thermodynamic formalism for weakly damped polaritons. We analyzed, in particular, the energy distribution, heat dissipation and maximum power as a function of exciton-phonon coupling in stationary state for a dissipative JCM. The formalism introduced in Ref. \citenum{ochoa2022quantum} overcomes the difficulties in propagating a system with slow dissipative/relaxation processes driven by a periodic-in-time external field. We showed that when the polartion-external field coupling is weak, there is an efficient and approximate form for the polariton's density-matrix propagator, which readily allows computing the polariton stationary state. Significantly, we found that when the driving strength is independent of the exciton-phonon coupling, the energy stored by the polariton, the thermodynamics performance and the irreversible heat are maximal in resonance.

In this paper we extend the study of the dissipative JCM model for lossy polaritons introduced in Ref.\ \cite{ochoa2022quantum} focusing in the dynamic and thermodynamic signatures resulting from the phonon anharmonicity and variations in the coupling between the polariton and the periodic driving field. Several studies\cite{campos2021generalization,campos2022generalization} report on the importance of accounting for anharmonicity in systems comprising molecules in cavities. Our results suggest that anharmonicity can limit the performance and stored energy in a periodically driven polariton, occasionally shifting from resonance to an off-resonance polariton state the configuration with the largest stored energy capacity upon driving.  We also find that the stationary polariton energy is larger when the driving field modulates the phonon degrees of freedom, but can be limited by saturation at the exciton. When the external field drives simultaneously the exciton and phonon, destructive interference occasionally occurs manifesting in a nontrivial polariton energy dependence on the external field-polariton coupling strength. A figure of merit defined in terms of the average work and irreversible heat, reveals that the phonon anharmonicity can improve the relative efficiency of the driving process while still reducing the absolute polariton energy. 

 The organization of the paper is as follows. In Sec.\ \ref{sec:model} we introduce the dissipative anharmonic JCM, our polariton model, describing dynamic and thermodynamic properties for generic periodic driving protocols. Then, in Sec.\ \ref{sec:energy}, we analyze the anharmonic signatures in the energy distribution and thermodynamics for polariton coupled to the external driving field to the exciton. Next, in Sec.\ \ref{sec:driving} we analyze the polariton's response to independent and concerted modulations of exciton and phonon state. We summarize in Sec.\ \ref{sec:conclusions}.


\section{Dissipative Jaynes-Cummings model dynamics and thermodynamics}\label{sec:model}

Our model for an excitonic polariton consists of a two-level exciton, with energies $\varepsilon_i$, and anharmonic phonon, with a characteristic energy $\omega$ and anharmonicity constant $\chi$, which are coupled with coupling strength $V$ ($\hbar = 1$)
\begin{align}
  \hat H_{\rm S} =&   \sum_{i=1}^2 \varepsilon_i \hat d_i^\dagger \hat d_i + \hat H_{\rm pho}(\omega,\chi) + V \hat d_2^\dagger \hat d_1 \hat a +V^* \hat a^\dagger \hat d_1^\dagger \hat d_2 \label{eq:Hsys}
\end{align}
In Eq.\ \eqref{eq:Hsys},  $\hat d_i^\dagger $ ($\hat d_i$) is the creation (annihilation) operator for an electron in the $i$-th level, and $\hat a^\dagger$ ($\hat a$) creates (annihilates) a phonon. $\hat H _{\rm pho}(\omega,\chi)$ is the Hamiltonian of the free phonon and, in the phonon-mode basis $\{ | m \rangle \}$, it is a diagonal operator with nonzero elements
\begin{equation}\label{eq:hpho}
 \langle m | \hat H _{\rm pho} | m \rangle = \omega m + \omega \chi (m +1/2)^2.
\end{equation}
We set the zero-point energy to zero in Eq.\ \eqref{eq:hpho}. Note that when $\chi$ vanishes,  $ \hat H _{\rm pho} \to \omega \hat a^\dagger \hat a$. The system Hamiltonian in Eq.\ \eqref{eq:Hsys} can be analytically diagonalized obtaining eigenvalues
\begin{multline}
  \label{eq:eigen}
  \lambda_{\pm}^{(m)}= \frac{\varepsilon_2+ \varepsilon_1 }{2}+\omega \left( \frac{1}{2}+ m \right)\\ - \omega \chi \left(\frac{5}{4}+m(m+2)\right)  \pm \Delta,
\end{multline}
with the discriminant $\Delta$ given by
\begin{equation}
  \label{eq:Delta}
  \Delta = \sqrt{|V|^2+ \left(\frac{\varepsilon_2 -\varepsilon_1 - \omega}{2}+ \omega \chi (1+m)  \right)^2}.
\end{equation}
The eigenvalues in Eq.\ \eqref{eq:eigen} converge to the standard JCM form when $\chi \to 0$ and, if the exciton and phonon characteristic energies are in resonance (i.e., $\varepsilon_2 -\varepsilon_1 = \omega$),  they assume the simpler form $\lambda_{\pm}^{(m)} \to \omega (m+1) \pm |V|$.

Exciton and phonon forming the polariton dissipate energy to their independent thermal baths  $\hat H_X = \sum_k \varepsilon_k \hat c_k^\dagger \hat c_k$ and  $\hat H_P =\sum_k \omega_k \hat b_k^\dagger \hat b_k$ according to the linear coupling
\begin{align}
   \hat H_{\rm IX} =& \sum_k W^X_{k} \hat c_k^\dagger \hat d_1 ^\dagger \hat d_2 + W^{X *}_{k} \, \hat d_2 ^\dagger \hat d_1 c_k \label{eq:HIX}\\
  \hat H_{\rm IP} =& \sum_k W^P_k \hat b_k^\dagger \hat a + \text{ h.c} \, \label{eq:HIP} ;
\end{align}
with coupling strengths $W^X_{k}$ and $W^P_{k}$.

The polariton is periodically modulated by an external field, with frequency $\omega'$, driving the exciton and phonon population following the generic driving Hamiltonian  
\begin{align}
  \hat H_{d}(t) =& 2 A_X \cos(\omega' t ) \notag\\
  &\hspace{0.8cm}\sum_{m} | 1, m \rangle \langle m, 2 | + | 2, m \rangle \langle m, 1 | \notag \\ 
  &+ 2 A_P \cos(\omega' t ) \notag \\
  &\hspace{0.8cm}\sum_{i\in \{1,2\}} \sum_{m} | i, m \rangle \langle m+1, i| + | i, m+1 \rangle \langle m, i |.\label{eq:Hd}                  
\end{align}
In Eq.\ \eqref{eq:Hd}, $A_X$ and $A_P$ represent the corresponding exciton and phonon driving strength energies -- which are proportional to the incident field amplitude and polariton response-- and $i,m$ are the electronic quantum number and phonon mode.

Following Ref.\ \cite{ochoa2022quantum}, we solve the the Liouville-von Neumann equation for the density matrix in the interaction picture with interacting Hamiltonian $\hat H_d(t)+ \hat H_I$,  invoking the Born-Markov approximation. The resulting time-local quantum master equation (QME)
\begin{align}
  \frac{d}{dt} \vec \rho(t) = -\left(\mathcal{D} + \mathcal{L}_X+\mathcal{L}_P +\mathcal{L}_d(t)\right) \vec \rho(t) \label{eq:rhot},
\end{align}
is valid to second order in $W^X, W^P$, $A_X$ and $A_P$; and holds for arbitrary exciton-phonon coupling strength $V$. The QME in Eq.\ \eqref{eq:rhot} propagates the polariton's reduced density matrix $\vec \rho(t)$, which we write as a vector with elements   $\rho_\alpha^\beta (t) = \langle \alpha | \rho(t) |\beta \rangle$ in the basis $\{ |\alpha \rangle\}$ of $\hat H_{\rm S}$ eigenstates. In Eq.\ \eqref{eq:rhot}, the operator $\mathcal{D}$ is diagonal with matrix element  $\mathcal{D}_{\alpha_2 \beta_2}^{\alpha_1 \beta_1}   = -i (\lambda_{\alpha_1} -\lambda_{\beta_1}) \delta_{\alpha_1}^{\alpha_2} \delta_{\beta_1}^{\beta_2}$ describing the free evolution of the polariton;  $\mathcal{L}_X$ and $\mathcal{L}_P$ are operators accounting for the independent exciton and phonon relaxation which are proportional to the damping rates $\Gamma_\square = 2 \pi \sum_k |W_k^\square|^2 \delta (\omega_k - \omega)$ with $\square = X, P$, respectively. Reference \cite{ochoa2022quantum} provides the explicit form of these operators. The external-field induced modulation leads to the time-dependent operator $\mathcal{L}_d(t)$, which we conveniently write as a linear combination of two operators $\mathcal{L}_{dX}(t)$ and $\mathcal{L}_{dP}(t)$. $\mathcal{L}_{dX}$ arises from the first term in Eq.\ \eqref{eq:Hd}, is proportional to $A_X^2$ and corresponds to the operator $\mathcal{L}_d(t)$ in Ref.\ \cite{ochoa2022quantum}. Likewise, $\mathcal{L}_{dP}$ is bounded by $A_P^2$ and accounts for the phonon external field modulation. 

Next, we solve the QME in Eq.\ \eqref{eq:rhot} by truncating the system size to include only the lowest $m_o$ phonon modes -- higher energy modes will not significantly contribute to the full dynamics due to the polariton environment-induced relaxation -- obtaining and approximate form for the dissipative time propagator for the density matrix
\begin{align}
  \vec \rho(t) = \exp\left[-(\mathcal{D} + \mathcal{L}_X+\mathcal{L}_P) t  - \int_0^t \mathcal{L}_d(t') dt' \right] \vec \rho(0),\label{eq:rhoExp}
\end{align}
 where $\rho(0)$ is the polariton initial equilibrium state. Utilizing the propagator in Eq.\ \eqref{eq:rhoExp}, we obtain the polariton stationary  state by considering a propagation time $t$ larger than the relaxation times given by $\Gamma_X$, and $\Gamma_P$.  In Appendix \ref{ap:solQME}, we show that the solution to Eq.\  \eqref{eq:rhot}, with the driving Hamiltonian $\hat H_d$ in Eq.\ \eqref{eq:Hd}, is local-in-time, and that Eq.\ \eqref{eq:rhoExp} is the approximate form valid whenever the damping rates and $A_X$, $A_P$ are small. In fact, we find that corrections to Eq.\ \eqref{eq:rhoExp} accounting for contributions from commutators of $\mathcal{L}_X$, $\mathcal{L}_P$, $\mathcal{L}_{dX}$ and $\mathcal{L}_{dP}$ are bounded by products of the damping rates and driving field strengths $\Gamma_X$, $\Gamma_P$, $A_X$ and $A_P$. These terms are potentially relevant in the transient regime.   

\begin{figure}[tb]
  \centering
  \includegraphics[scale=0.5]{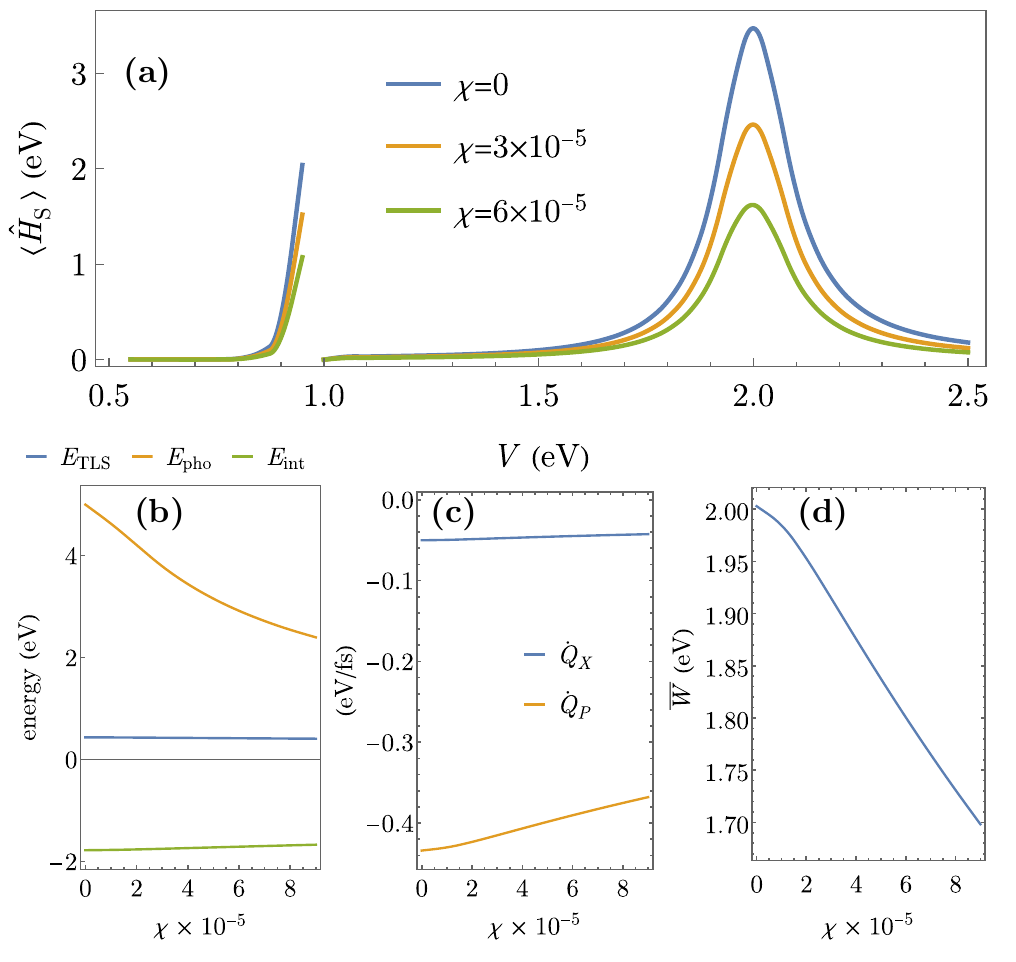}
  \caption{Anharmonic polariton energy distribution. (a) Stationary anharmonic polariton energy $\langle \hat H_S \rangle$ as a function of coupling strength $V$ for three different values in the anharmonicity constant $\chi$, $0$ (blue), $3 \times 10^{-5}$ (orange), $6 \times 10 ^{-5}$ (green). (b) Stationary exciton, phonon and interaction energy $E_{\rm TLS}$, $E_{\rm pho}$ and $E_{\rm int}$ (b) heat rates $\dot Q_X$ and $\dot Q_P$, and (d) total work  $\overline{W}$ at resonance $V = 2\omega$ as a function of the anharmonicity parameter $\chi$. Parameters for this system are $\omega'=\omega=1.0$ eV, $\Gamma_X=0.2$ eV, $\Gamma_P=0.4$ eV, $A_X = 0.1$ eV, $A_P= 0$, $\varepsilon_1 =0$, $\varepsilon_2 = 1.0$ eV, $T= 300$ K. Propagation time $t=8.27$ ps.\label{fig:EneAnh}}
\end{figure}

 The polariton thermodynamic properties in stationary state follow from its density matrix and the definitions for heat and work\cite{ochoa2022quantum},
\begin{align}
  \dot Q_\square(t) =& - {\rm Tr} \left[ L_\square(t) \hat H_{\rm S} \right ] \label{eq:dotQ}\\
\dot W_\square(t) =& - {\rm Tr} \left [ L_{d\square} (t) \hat H_{\rm S} \right ]. \label{eq:dotW}  \hspace{0.5cm}(\square = X, P),
\end{align}
where
\begin{align}
  L^{\;\;\beta_1}_{\square\, \alpha_1} = \sum_{\alpha_2 \beta_2}\mathcal{L}^{\;\;\; \alpha_1 \beta_1}_{\square\, \alpha_2 \beta_2} \rho^{\beta_2}_{\alpha_2}(t)  \hspace{0.5cm}(\square = X, P , dX, dP),
\end{align}
From the polariton's von Neumann entropy 
\begin{align}
   S(t) = & - {\rm Tr } \left[ \rho(t) \ln \rho(t) \right],\label{eq:Stot}
\end{align}
we obtain the entropy change rate 
\begin{align}
  \dot S(t) =& \sum_{\square=X,P}\dot S_\square(t)+\dot S_{d\square}(t), \\
  \dot S_{\square} =&  {\rm Tr } \left[ L_\square(t) \ln \rho(t) \right] \hspace{0.5cm} (\square = X, P, dX, dP) \label{eq:dotS},
\end{align}
which as a consequence of Spohn's theorem\cite{spohn1978entropy,alicki1979quantum} provides the definition of irreversible heat rate,
\begin{align}
  \dot Q_{\rm irrev}(t) =& \beta_o^{-1} Tr[{(\mathcal{L}_X(t)+\mathcal{L}_P(t)) (\ln \rho_o - \ln \rho(t)\,)]} \notag\\
  =& \beta_o^{-1}(\dot S_X(t) + \dot S_p(t)) - (\dot Q_X(t) + \dot Q_P(t)), \label{eq:Qirrev}
\end{align}
relative to the polaritons equilibrium density matrix  $\rho_o = e^{-\beta_o \hat H_{\rm S}}/{\rm Tr}[e^{-\beta_o \hat H_{\rm S}}]$, with $\beta_o = 1/(k_B T_{\rm env})$ and $T_{\rm env}$ the absolute temperature.

\section{Anharmonic polaritons}\label{sec:energy}

Next, we numerically investigate the polariton energy distribution between the exciton, phonon and interaction parts and how such distribution depends on the polariton's anharmonic character and exciton-phonon coupling strength. We also analyze how energy dissipation depends on these factors. For simplicity, we consider the case where the driving field directly modulates the exciton by setting $A_P = 0$ postponing until the next section the study of arbitrary external periodic drivings.  

\begin{figure}[th]
  \centering
  \includegraphics[scale=0.6]{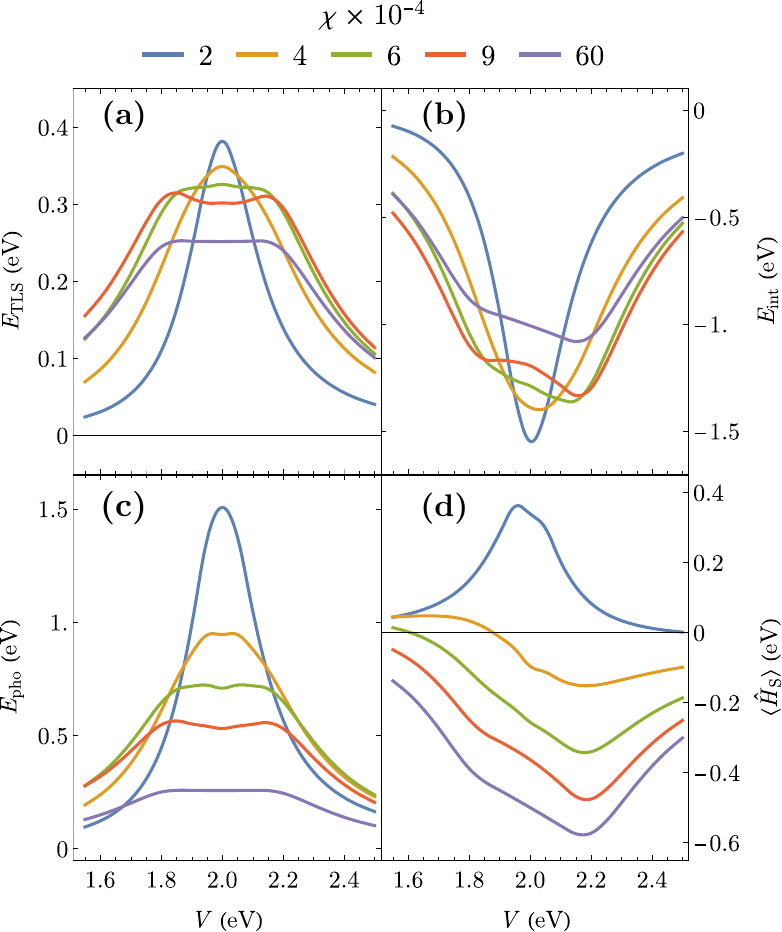}
  \caption{Energy distribution in strongly anharmonic polaritons. (a) Exciton energy $E_{\rm TLS}$ (b) exciton-phonon interaction energy $E_{\rm int}$, (c) phonon energy $E_{\rm pho}$, and (d) total energy $\langle \hat H_{\rm S} \rangle$  as a function of the exciton-phonon coupling strength $V$. The lineshapes shown are for polaritons with anharmonicity constant $\chi$ varying from  $2 \times 10^{-4}$ to $6 \times 10^{-3}$. Other paremeters are as in Fig.\ \ref{fig:EneAnh}. \label{fig:StrongAnh}}
\end{figure}

For the numerical analysis, we study a polariton in resonance, with  $\omega= \varepsilon_2= 1$ eV and $\varepsilon=0$, taking as the initial state the exciton in the ground state. The damping rates are assumed to be frequency-independent, but we remark that the model does not depend on this assumption. The  incident field frequency $\omega'$ is set to match the phonon frequency $\omega$, as well as the exciton transition energy. This choice on the exciton energy provides a direct reading on the exciton population from the reduced energy of the system, whenever we can interpret $E_{\rm TLS} = \varepsilon_2 \langle  d_2^\dagger d_2 \rangle$ as $\rho_{22}$. We will also consider the phonon energy $E_{\rm pho} = \langle \hat H_{\rm pho} (\omega, \chi) \rangle$ and $E_{\rm int} = V \langle \hat d_2^\dagger \hat d_1 \hat a + \hat a^\dagger \hat d_1^\dagger \hat d_2 \rangle $.  In the harmonic limit as discussed in Ref.\ \citenum{ochoa2022quantum}, the linear dependence on $V$ of the JCM eigenvalues and the $\hat H_d(t)$ form imply maximum energy absorption populating higher hybrid energy states when $V=2 \omega$. This corresponds to the second type of resonance: the one between the driving field and the hybrid state transition $\varepsilon_1 \to \lambda_{-}^{0}$.

We begin by studying the hybrid exciton-phonon total energy profile, as a function of the coupling strength $V$, and its deviation from the harmonic limit. Figure \ref{fig:EneAnh}(a) shows $\langle \hat H_S \rangle$ for three polaritons characterized by different values in the anharmonicity constant $\chi$.  The system with $\chi=0$ is the harmonic model analyzed in Ref.\ \citenum{ochoa2022quantum}. We note that while the qualitative dependence of the system energy $\langle H_S \rangle$ on $V$ is preserved in weakly anharmonic polaritons with $\chi < 10^{-4}$, the total energy is consistently lower as the anharmonic character increases. Indeed, as illustrated in Fig. \ref{fig:EneAnh}(b) for a polariton in resonance $V=2\omega$, with $\omega' =\omega$, the phonon energy $E_{\rm pho}$ for a system with  $\chi = 9 \times 10^{-5}$ is about 50\% of its expected value in the harmonic limit $\chi =0$, while the stationary $E_{\rm int}$ and $E_{\rm TLS}$ do not vary significantly. Thus, the observed energy difference is the result of the lower energy storage capacity in the phonon modes. Figure \ref{fig:EneAnh}(c) shows the reversible heat rates reveling a linear decrease in the absolute value of both  $\dot{Q}_P$  and $\dot{Q}_X$ with $\chi$. Similarly, the mean work rate per period, $\overline{W}=\int_{\tau N}^{\tau (N+1)}\dot{W(t)} dt$, in stationary state $\overline{W}$ decreases with an approximate linear dependence on $\chi$, as illustrated in Fig.\ \ref{fig:EneAnh}(d), suggesting that even in the situation in which the driving field couples exclusively with the exciton, the phonon anharmonicity can modify the maximum power during periodic modulation.

\begin{figure}[t]
  \centering
  \includegraphics[scale=0.5]{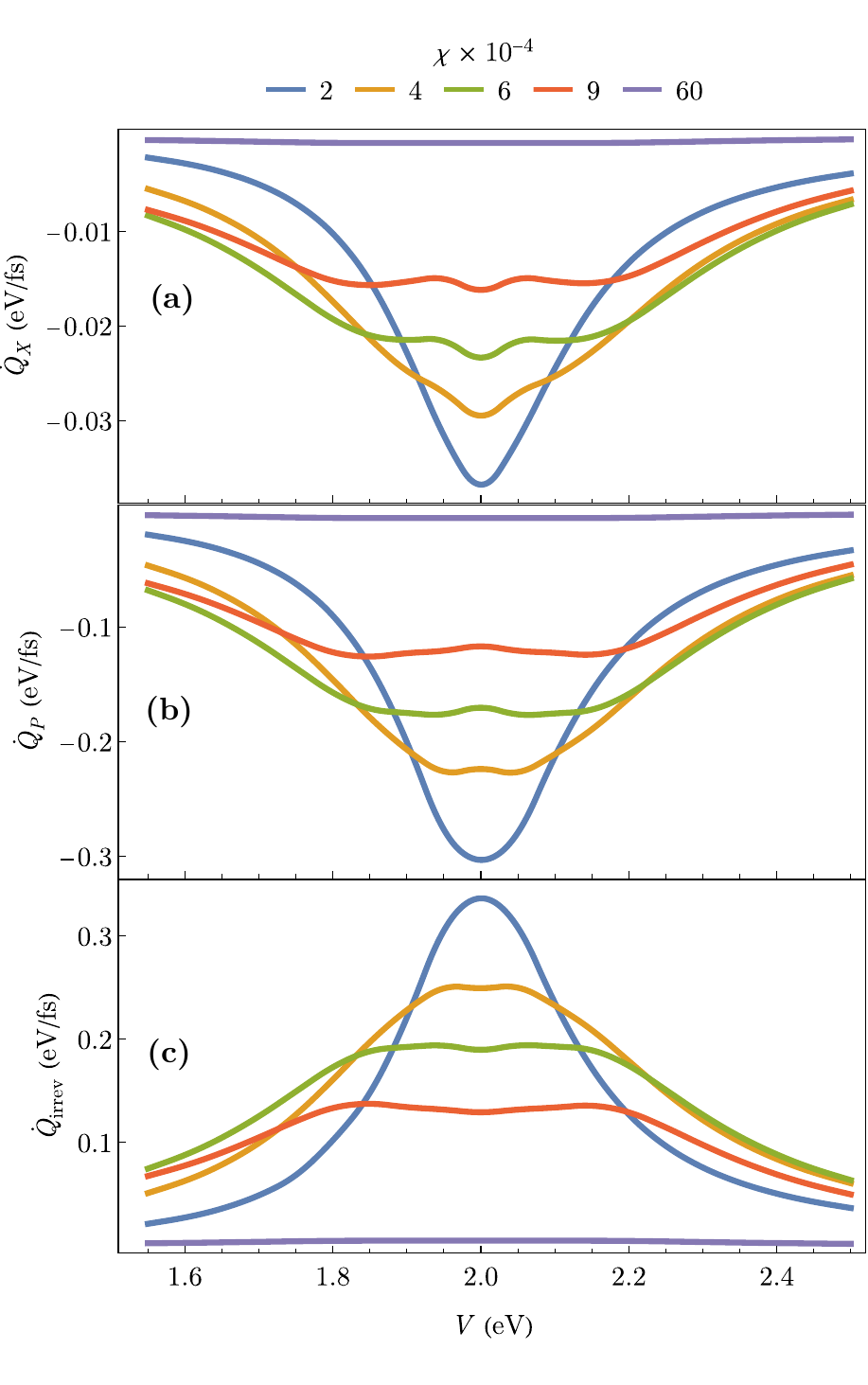}
  \caption{Reversible and irreversible heat rates in a periodically-driven strongly-anharmonic polariton. (a) $\dot Q_X$  (b)$\dot Q_P$ (c) $\dot Q_{\rm irrev}$ . The lineshapes shown are for polaritons with anharmonicity constant $\chi$ varying from  $2 \times 10^{-4}$ to $6 \times 10^{-3}$. Other parameters are as in Fig.\ \ref{fig:EneAnh}.\label{fig:HeatAnh}}
\end{figure}

The functional dependence of the polariton energy on $V$ takes a remarkably different form from the one observed near the harmonic limit when the polariton's anharmonicity character is stronger. For the particular system under consideration, we observe such transition in Fig.\ \ref{fig:StrongAnh} occurring when $\chi > 10^{-4}$. In constrast to the finding for weakly anharmonic systems, for $\chi$ between $10^{-4}$ and $10^{-3}$, we observe in Figs.\ \ref{fig:StrongAnh}(a) and \ref{fig:StrongAnh}(b) a decrease on the absolute exciton $E_{\rm TLS}$ and the interaction $E_{\rm int}$ energies near resonance. Moreover, the lineshape broadens for larger values in $\chi$, deviating from the Lorentzian form identified for the harmonic limit in Ref.\ \cite{ochoa2022quantum}, resulting incidentally in larger absolute exciton and interaction energies off-resonance. Significantly, the resonance energy varies from a highly excited state, close to the population saturation in the harmonic limit to one third population ratio, i.e., $\rho_{22}/\rho_{11} \sim 0.25/0.75$. For an exciton decoupled to the phonon mode and under similar driving conditions, the weak damping rate allows for the exciton energy to reach values close to 0.5 eV corresponding to an excited population of $\rho_2 \sim 0.5$, i.e., near saturation. We conclude that anharmonicity character is detrimental near resonance but can enhance the energy absorption at the exciton off-resonance (i.e., $V \neq 2 \omega$). The phonon energy, presented in Fig.\ \ref{fig:StrongAnh}(c), follows a similar trend to the one observed for the exciton with a more considerable reduction in maximum energy with $\chi$. Consequently, we find in Fig.\ \ref{fig:StrongAnh}(d) that the total polariton energy is dominated by the interaction energy, leading to negative energies relative to the free exction ground state with a minimum in polariton energy occurring off-resonance $V > 2 \omega$.

The functional dependence of reversible and irreversible heat rates on $V$  also changes with the polariton anharmonic character as illustrated in Fig.\ \ref{fig:HeatAnh}. The trends are similar to the ones described above for the phonon or exciton energies: broadening of the lineshape near resonance with the consequent decay the the absolute value of these rates at $V = 2 \omega$ with a relative increase off resonance. Remarkably, for $\chi > 10^{-3}$, $E_{\rm pho}$, $\dot Q_X$, $\dot Q_P$ and $\dot Q_{\rm irrev}$ drop several orders of magnitude.

Finally, in Appendix \ref{ap:weakGamma}, we show the how the polariton energy changes for weaker damping rates than those consider in Figs.\ \ref{fig:StrongAnh}, revealing that for polaritons with very weak environmental relaxation high energy absorption are possible in a large interval of coupling strengths near the resonance condition $V= 2 \omega$.




\begin{figure}[th]
  \centering
  \includegraphics[scale=0.42]{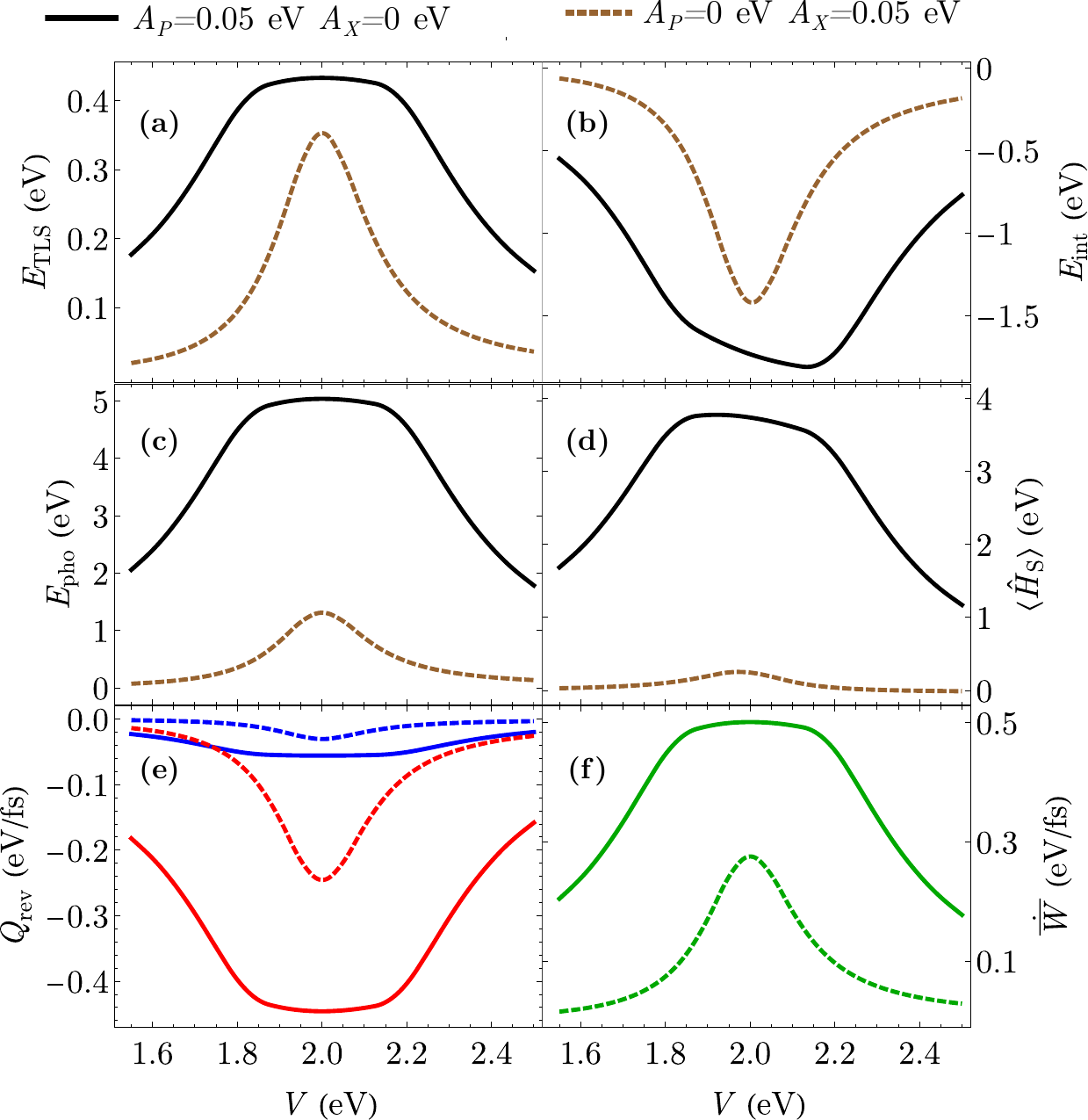}
  \caption{Comparison in the energy distribution, dissipated heat and mean work performed on a periodically driven polariton when the external field couples to the exciton ($A_X = 0.05$ eV and $A_P=0 $, dashed lines) versus coupling to the phonon($A_X = 0$ and $A_P=0.05$ eV, solid lines). (a) exciton energy $E_{\rm TLS}$ (b) interaction energy $E_{\rm int}$ (c) phonon energy $E_{\rm pho}$ (d) total energy $\langle \hat H_{\rm S} \rangle$, (e) reversible heat rates $\dot Q_{X}$ (blue), and $\dot Q_{P}$ (red), and (f) mean work rate $\dot{\overline{W}}$ . The polariton is harmonic with $\chi = 0$. Other parameters are as in Fig.\ \ref{fig:EneAnh} .\label{fig:AxvsAP}}
\end{figure}

\section{ Exciton, phonon, and exciton-phonon driving}\label{sec:driving}

In this section we study general driving regimes that include direct coupling of the external driving field to the phonon, which are expected to be relevant in, for instance, plasmonic cavities.

We begin by comparing the two extreme cases in which the driving field modulates exclusively the exciton and the phonon forming the harmonic polariton ($\chi = 0$) in Fig.\ \ref{fig:AxvsAP}, where we juxtapose the stationary state for a periodically driven polariton in both cases. We take as the external driving coupling strengths  $A_X = 0.05$ eV and $A_P = 0$ in one case, and $A_X=0$ and $A_P=0.05$ eV in the other one. All other parameters are identical, such that any observed dissimilarity results from differences in the driving protocols. In Fig.\ \ref{fig:AxvsAP}(a) we find that $E_{\rm TLS}$ is consistently smaller when the driving field modulates exclusively the exciton population for all values in $V$. Moreover, when the external field modulates the phonon, the exciton energy is nearly the saturation energy for an interval of $V$ values close to the resonance coupling strength $V=2\omega$, displaying a plateau in energy. Put it other way, this observation indicates that the exciton excited population is closed to the maximum allowed by the weak system interaction with the heat reservoir, suggesting also the lack of population inversion ($E_{TLS} < 0.5$ such that $\rho_2 < \rho_1 $). The exciton-phonon coupling energy is also larger in absolute value when the external field couples to the phonon (see Fig.\ \ref{fig:AxvsAP}(b)). The phonon energy in Fig.\ \ref{fig:AxvsAP}(c) is also larger in this case, showing a plateau near resonance similar in form to the one observed for $E_{\rm TLS}$. Surprisingly, this finding reveals that $E_{\rm pho}$ can be limited by the maximum excited population achievable by the exciton when the exciton and phonon interact strongly. The total polariton energy, \ref{fig:AxvsAP}(d), is about one order of magnitude larger when the field couples to the phonon only, and shows an asymmetry near resonance resulting from the linear dependence on $V$ of $E_{\rm int}$. We conclude that energy storage in the polariton is enhanced when the external field couples to the phonon in all cases. Figure \ref{fig:AxvsAP}(e) shows $\dot Q_X$ and $\dot Q_P$ revealing that these rates follow similar trends to those observed for $E_{\rm TLS}$ and $E_{\rm pho}$, as described above. Lastly, we find in Fig.\ \ref{fig:AxvsAP}(f) that the periodic driving of the phonon results in larger mean power $\dot{\overline{W}}$.

\begin{figure}[t]
  \centering
  \includegraphics[scale=0.42]{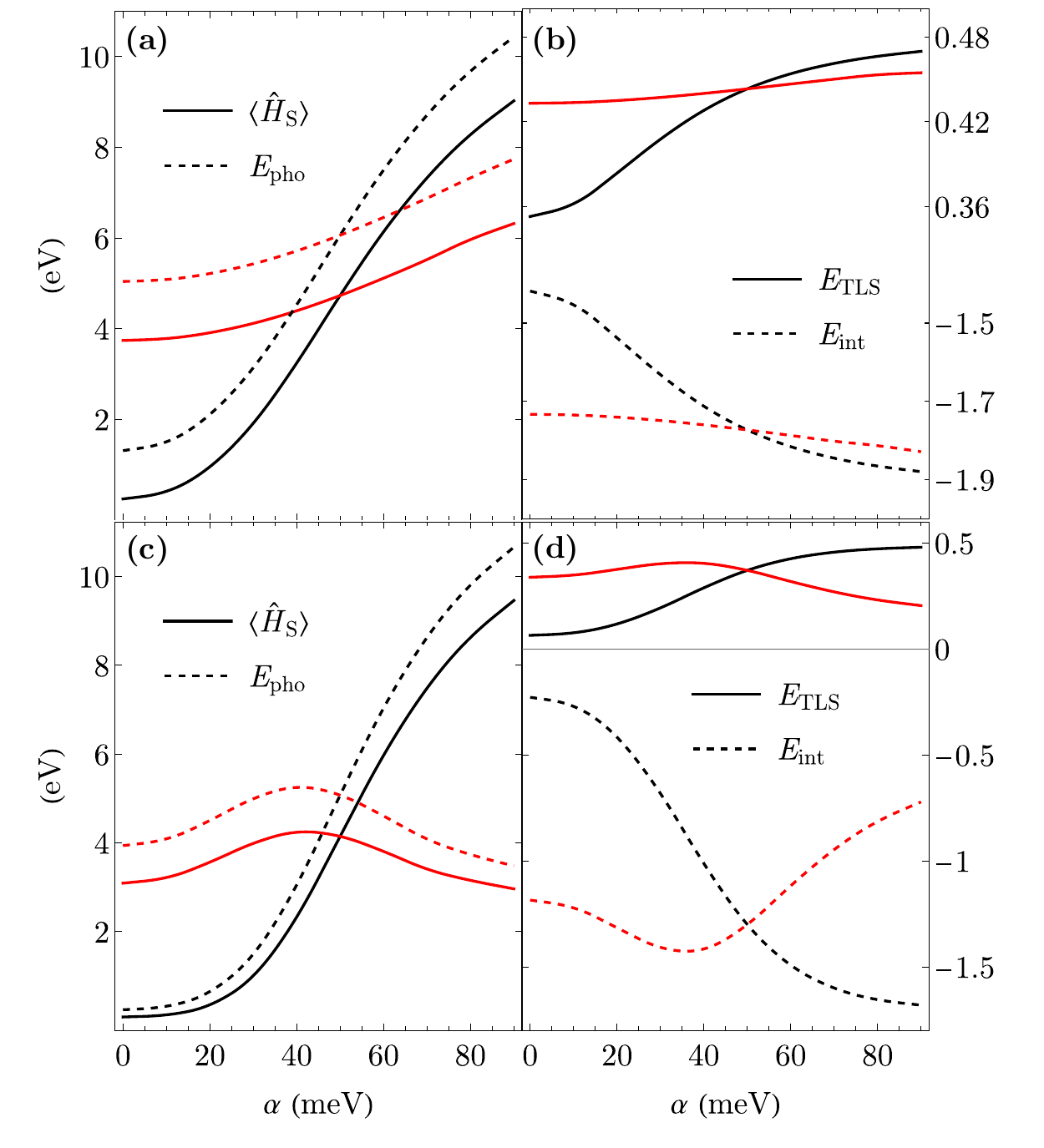}
  \caption{Simultaneous exciton and phonon driving. The black lines are for a protocol with fixed $A_X = 0.05$ eV and varying $A_P$, such that $A_P= \alpha$. The read lines show the opposite situation, with $A_X = \alpha$ and fixed $A_P= 0.05$. Panels (a) and (b) show energies in resonance, i.e., $V=2 \omega$, while  (c) and (d) are off resonance with $V= 1.75\, \omega$. Panels (a) and (c) show total and phonon energy, $\langle \hat H_{\rm TLS} \rangle$ and $E_{\rm pho}$; (b) and (d) show the exciton and interaction energies, $E_{\rm TLS}$ and $E_{\rm pho}$, for each protocol. The polariton is harmonic with $\chi = 0$ in all cases. Other parameters are as in Fig.\ \ref{fig:EneAnh}.  \label{fig:BothA}}
\end{figure}

Next, we analyze mixed situations in Fig.\ \ref{fig:BothA}, where the external driving field couples simultaneously to both the exciton and the phonon with varying strengths $A_X$, $A_P$. Specifically, we consider how the total energy and its distribution changes when we fix either $A_X$ or $A_P$ to $0.05$ eV while modulating the other coupling parameter from $0$ to $0.09$ eV. Figures \ref{fig:BothA}(a) and (b) present results for a polariton in resonance ($V=2 \omega$). In this case, the total energy stored by the polariton is considerably enhanced by several orders of magnitude when $A_P$ increases in the range considered while we fixed $A_X$. In contrast, the inverted situation leads to an increase of nearly 50\% and minor variations in $E_{\rm TLS}$ and $E_{\rm int}$. These patterns may change off-resonance. As an illustration, in Figs.\ \ref{fig:BothA}(c) and (d) we investigate the same protocols when $V = 1.75 \omega$, finding a nonmonotonic change for fixed $A_P$ and varying $A_X$. Looking at the variation in $E_{\rm TLS}$, $E_{\rm pho}$, and $\langle \hat H_{\rm S} \rangle$, we observe that each one of these energies reaches its maximum value near $A_X = \alpha= 0.04$ eV, resulting in a drop in the same as $A_X$ increases for $A_X > 0.04$ eV. This observation reveals the occurrence of destructive interference between the exciton and photon dynamics induced by the external field for particular coupling strenghts during their simultaneous periodic driving.

\begin{figure}[t]
  \centering
  \includegraphics[scale=0.36]{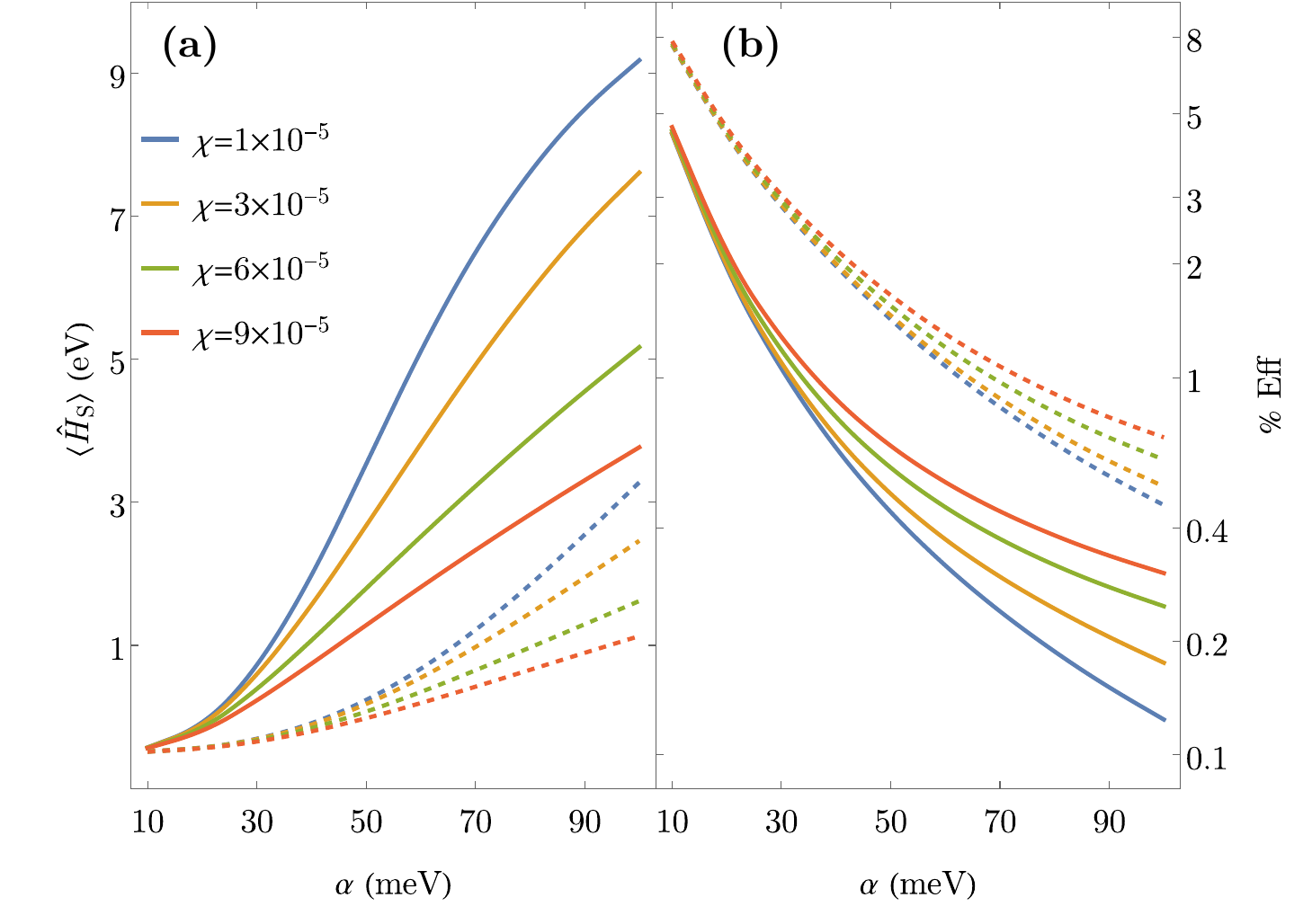}
  \caption{Total stored energy $\langle \hat H_{\rm S} \rangle$ (a) and driving efficiency \%Eff Eq.\ \eqref{eq:figmer} (b) for the periodic modulation of several anharmonic polaritons as a function of driving strengths $A_X$ and $A_P$. We show results for two driving forms: fixed $A_X = 0$ and varying $A_P = \alpha$ (Solid lines), and varying $A_X = \alpha$ with fix $A_P = 0$ (dashed lines). Other parameters are as in Fig. \ref{fig:EneAnh}.  \label{fig:ModratioEff}}
\end{figure}

Finally, we return to our investigation of anharmonic polaritons in Fig.\ \ref{fig:ModratioEff}, where we show stationary state properties for several anharmonic polaritons periodically modulated according to the two protocols discussed in Fig.\ \ref{fig:AxvsAP}. In this case we limit our consideration to polaritons in resonance, and find that for both protocols the total energy $\langle \hat H_{\rm S} \rangle$  decreases with $\chi$, and increases as the coupling term $A_X$ or $A_P$ (Fig.\ \ref{fig:ModratioEff}(a)). Moreover, the trend identified above, indicating that the energy stored in the polariton is larger when the external field modulates the phonon population persists in weakly anharmonic polaritons.  We introduce the following figure of merit to evaluate the efficiency driving protocols
\begin{equation}
  \label{eq:figmer}
   {\rm Eff} = \frac{\bar W - \bar Q_{\rm irr}}{\bar W},
\end{equation}
where $\bar W$ and $\bar Q_{\rm irr}$ are the mean work and irreversible heat per period. The figure of merit in Eq.\ \eqref{eq:figmer} is the ratio between the total energy transferred from the external field to the polariton and the actual ``useful'' energy in stationary state. When we evaluate Eff in  Fig.\ \ref{fig:ModratioEff}(b), we observe higher efficiencies the larger $\chi$ is, in both protocols.  In addition, the figure of merit in Eq.\ \eqref{eq:figmer} quickly decays with the external field coupling strength. Surprisingly, the periodic modulation of the exciton population has consistently  larger values in the figure of merit than the protocol driving the phonon. Based on these results, we conclude that while an external periodic field coupled to phonon favors larger energy storage in the polariton, the ratio between input energy and useful energy per period of time is smaller when compared with the direct modulation of exciton population.

\section{Conclusions}\label{sec:conclusions}
We presented a systematic study of the dynamics and thermodynamic properties of anharmonic polaritons under several periodic driving protocols. When the phonon forming the polariton is weakly anharmonic, the energy distribution and thermodynamic properties are very similar to those observed in harmonic polaritons\cite{ochoa2022quantum}, modifying almost exclusively the energy stored in the phonon modes. Still, when the anharmonic character is moderate or significant, the hybrid system's total energy, exciton, and interaction energies drop. In particular, moderate anharmonicity lowers the exciton energy below saturation otherwise expected for a weakly damped, uncoupled exciton, also observed close to resonance in the harmonic limit. In this sense, anharmonicity can be detrimental. We also note destructive interference during the simultaneous exciton and phonon driving at a particular off-resonance driving frequency, resulting in lower exciton and phonon energies than those observed for the uncoupled components. Moreover, after defining a figure of merit for the polariton periodic modulation in terms of the ratio between the mean work and useful energy,  we found that while an external periodic field coupled to the phonon enhances the polariton's energy storage capacity, the ratio input energy/useful energy per period is smaller than when the external field modulates only phonon population.
 The results presented here allow us to understand the polariton response when the energy spectrum for the hybrid system deviates from the standard Jaynes-Cumming model and, in this regard, can be extended to study more complex situations where ultra-strong coupling emerges. Future studies will address this case and investigate additional model properties, such as the role of the spectral density and contributions resulting from the correlations between baths.

\appendix
\section{Propagator}\label{ap:solQME}
In this section we demonstrate that under weak periodic driving, the propagation of an intial state is well described by the local-in-time propagator in Eq.\ \eqref{eq:rhoExp}.

First, we note that the QME in Eq.\ \eqref{eq:rhot} can be written in the following form 
\begin{equation}
  \label{eq:diffA}
  \frac{d}{d t}\vec{\rho}(t) =- \left\{ \mathcal{A}_o+f_+(t)\mathcal{A}_++f_-(t)\mathcal{A}_-\right\}\vec{\rho(t)},
\end{equation}
where $\mathcal{A}_o$ and $\mathcal{A}_\pm$ are time-independent matrices, and $f_\pm(t)=2 \cos(\omega t )e^{\pm i\omega t} $. By direct integration we find
\begin{equation}
  \label{eq:rho1}
  \vec{\rho}(t)=\vec{\rho}(0)-\int_0^t dt_1  \left\{ \mathcal{A}_o+f_+(t_1)\mathcal{A}_++f_-(t_1)\mathcal{A}_-\right\} \vec{\rho}(t_1)
\end{equation}
which leads to the following form for the solution to the differential equation Eq.\ \eqref{eq:diffA}
\begin{equation}
  \label{eq:solA}
  \vec{\rho}(t)=\sum_{n=0}^\infty(-1)^n I_n(t) \vec{\rho}(0).
\end{equation}
with
\begin{align}
  I_0(t) =& 1 \label{eq:I1}\\
  \label{eq:In}
  I_n(t) =& \int_0^t dt_1 \left\{ \mathcal{A}_o+f_+(t_1)\mathcal{A}_++f_-(t_1)\mathcal{A}_-\right\} I_{n-1}(t_1).
\end{align}
We prove that Eq.\ \eqref{eq:solA} is indeed a solution to the QME, showing first that $\{ I_n(t) \}$ is a uniformly convergent series. Moreover, we show that Eq.\ \eqref{eq:solA} is time-local since we can evaluate the integrals defining $I_n(t)$ directly.

We begin by rewriting $I_n$ expanding the recursive series in Eq. \eqref{eq:In} 
\begin{align}
  I_n(t) =& \int_0^t dt_1 \int_0^{t_1}dt_2 \dots \int_0^{t_{n-1}}dt_n\notag\\
          &\hspace{0.3cm}\prod_{k=1}^n\left\{ \mathcal{A}_o+f_+(t_k)\mathcal{A}_++f_-(t_k)\mathcal{A}_-\right\}\\
  =& \sum_{l \in 3^n} \left(\prod_{k=1}^n  \mathcal{A}_o^{\delta_0^{l_k}} \mathcal{A}_-^{\delta_-^{l_k}}\mathcal{A}_+^{\delta_+^{l_k}}  \right)\times \notag\\
          &\hspace{0.7cm} \int_0^t dt_1 f_+(t_1)^{\delta_+^{l_1}}f_-(t_1)^{\delta_-^{l_1}}\notag \\
  &\hspace{1.4cm}\dots\int_0^{t_n-1} dt_1 f_+(t_n)^{\delta_+^{l_n}}f_-(t_n)^{\delta_-^{l_n}} \label{eq:In2}
\end{align}
In Eq. \eqref{eq:In2}, $l$ is an $n$-tuple $\{l_1,\dots,l_n\}$ with $\l_i \in \{ 0,1,2 \}$, representing also a single integer in $3^n$. The symbol $\delta_\pm^{l_i}$ is the Kronecker delta.  With the correspondence $1 \iff -$, $2 \iff +$, each $l \in 3^n$ represents one term resulting from the expansion of the product in Eq.\ \eqref{eq:In2}. We proceed to evaluate the integrals in Eq.\ \eqref{eq:In2} defining
\begin{equation}
  \label{eq:Fint}
  F(t,l_i)= \int_0^t dt'  f_+(t')^{\delta_+^{l_i}}f_-(t')^{\delta_-^{l_i}}, 
\end{equation}
such that
\begin{equation}
  \label{eq:Fint}
  \frac{d}{dt} F(t,l_i)=  f_+(t)^{\delta_+^{l_i}}f_-(t)^{\delta_-^{l_i}},
\end{equation}
and
\begin{align}
I_n(t) =& \sum_{l \in 3^n} \left(\prod_{k=1}^n  \mathcal{A}_o^{\delta_0^{l_k}} \mathcal{A}_-^{\delta_-^{l_k}}\mathcal{A}_+^{\delta_+^{l_k}}  \right)\times \notag\\
&\hspace{0.3cm} \int_0^t dt_1  \frac{d}{dt_1} F(t_1,l_1)\dots\int_0^{t_n-1} dt_1  \frac{d}{dt_n} F(t_n,l_n). \label{eq:In3}
\end{align}
Then we integrate by parts each integral in Eq.\ \eqref{eq:In3} utilizing the identity
\begin{equation}
 \frac{t_1^{n-1}}{(n-1)!} = \int_0^{t_1}dt_2 \dots \int_0^{t_{n-1}} dt_n.
\end{equation}
One finds, for instance, that
\begin{align}
  \int_0^t dt_1 \frac{d}{dt}& F(t_1,l_1) \int_0^{t_1}dt_2 \dots \int_0^{t_{n-1}} dt_n =\notag\\ & F(t,l_1)\frac{t^{n-1}}{(n-1)!}-\int_0^t dt_1F(t_1,l_1) \frac{t_1^{n-1}}{(n-1)!},\label{eq:int1}
\end{align}
for the case $l_1 \neq 0$ and $l_k = 0$ for $k > 0$.  Also note that there are $2 n $ integrals of this type, corresponding to $l$ with only one nonzero entry. All of these integrals lead to the same form on the first term in the right hand side of Eq.\ \eqref{eq:int1}.  More generally, for $l \in 3^n$, we have
\begin{align}
  \int_0^t dt_1  \frac{d}{dt_1}& F(t_1,l_1)\dots\int_0^{t_n-1} dt_1  \frac{d}{dt_n} F(t_n,l_n) =\notag\\
& \frac{t^{n-z_l}}{(n-z_l)!} \prod_{i \in n} F(t,l_i)+\mathcal{I}_n(t,l) \label{eq:IntFn}
\end{align}
where $z_l$ is the number of nonzero entries in $l$, and $\mathcal{I}_n(t,l)$ is the set of additional integrals resulting from integration by parts. We remark that the uniform convergence of the series follows form the factors  $t^{n-z_l}/(n-z_l)!$.  There are $n!/(n-z_l)!$ multidimentional integrals similar to the left-hand side in Eq.\ \eqref{eq:IntFn}, that corresponds to permutations of elements of $l$. These correspondance between integrals allows us to factor terms that are proportional to commutators between matrices $\mathcal{A}_o, \mathcal{A}$, and $ \mathcal{A}_+$. We do not attempt to calculate explicitly those terms here, but note that they are bounded by products of the intensity of the external field and damping rates.  Thus, we must split the integral in Eq.\ \eqref{eq:IntFn} into $n!/(n-z_l)!$ equivalent integrals, such that $n!/(n-z_l)!-1$ of them are factorized in terms proportional to the matrix commutators.  We obtain the exponential form for the propagator from the first term in the remaining integral
\begin{align}
  \frac{(n-z_l)!}{n!}\int_0^t dt_1  \frac{d}{dt_1}& F(t_1,l_1)\dots\int_0^{t_n-1} dt_1  \frac{d}{dt_n} F(t_n,l_n) =\notag\\
& \frac{t^{n-z_l}}{n!} \prod_{i \in n} F(t,l_i)+  \frac{(n-z_l)!}{n!} \mathcal{I}_n(t,l), \label{eq:IntFn2}
\end{align}
as follows
\begin{align}
  I_n(t) \approx& \sum_{l \in 3^n} \left(\prod_{k=1}^n  \mathcal{A}_o^{\delta_0^{l_k}} \mathcal{A}_-^{\delta_-^{l_k}}\mathcal{A}_+^{\delta_+^{l_k}}  \right) \frac{t^{n-z_l}}{n!} \prod_{i \in n} F(t,l_i)\\
  =&\frac{1}{n!} \sum_{l \in 3^n} \left(\prod_{k=1}^n  \mathcal{A}_o^{\delta_0^{l_k}} \mathcal{A}_-^{\delta_-^{l_k}}\mathcal{A}_+^{\delta_+^{l_k}}  \right)\times\notag\\
  &\hspace{3cm}t^{\delta_0^{l_k}} \int_0^t dt'  f_+(t')^{\delta_+^{l_k}}f_-(t')^{\delta_-^{l_k}}\\
  =&\frac{1}{n!}\left(\mathcal{A}_o t + \mathcal{A}_+ \int_o^t dt f_+(t)+\mathcal{A}_- \int_o^t dt f_-(t) \right)^n.\label{eq:In4} 
\end{align}
Finally, we replace Eq.\ \eqref{eq:In4} in Eq.\ \eqref{eq:solA} to obtain the approximate exponential form for the propagator.

\begin{figure}[ht]
  \centering
  \includegraphics[scale=0.42]{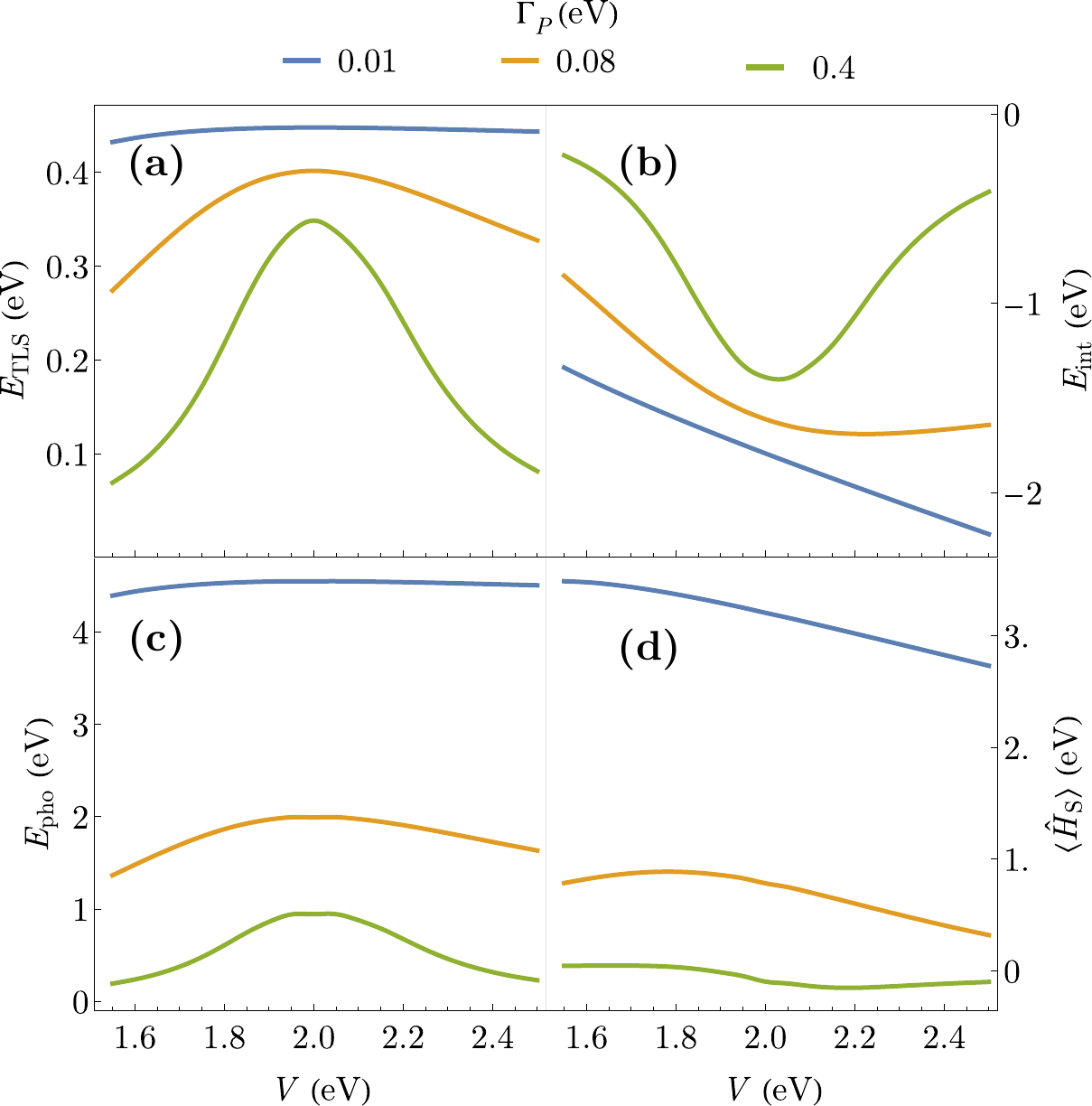}
  \caption{Energy distribution in a periodically-driven strongly-anharmonic polariton ($\chi = 4 \times 10^{-4}$ )as a function of the exciton-phonon coupling energy $V$, and for three different phonon relaxation rates $\Gamma_P = 0.01$ (blue), $0.08$ (orange), and  0.4 (green) eV. In each case $\Gamma_X = \Gamma_P/2$, and the propagation time was 827 ps. (a) exciton energy $E_{\rm TLS}$  (b) interaction energy $E_{\rm int}$ (c) phonon energy $E_{\rm pho}$ and(d) total energy $\langle \hat H_{\rm S} \rangle$. Other parameters are as in Fig. \ref{fig:EneAnh}. \label{fig:EneA}}
  
\end{figure}

\section{Energy distribution in weakly-damped anharmonic polaritions}\label{ap:weakGamma}

In this appendix we study how the energy distribution in a strongly anharmonic polariton under periodic driving varies as a function of damping rates. Figure \ref{fig:EneA}, presents $E_{\rm TLS}$, $E_{\rm pho}$, $E_{\rm int}$ and $\langle \hat H_S \rangle$ for polaritons with damping rates weaker than those considered in the main text, assuming an anharmonicity coefficient $\chi = 4 \times 10^{-4}$. We observe that, despite the relatively strong anharmonicity in the system, lower damping rates allow for larger off-resonance energy absorption, with exciton energies near saturation when $\Gamma_P = 2 \Gamma_X = 0.01 $ eV.




\bibliography{PolaritonReferences.bib}

\end{document}